\documentclass[reprint,aps,prx,balancelastpage,a4paper,floatfix,superscriptaddress]{revtex4-2}

\renewcommand \thesection{\arabic{section}}
\usepackage{titlesec}
\titlespacing*{\subsection}{0pt}{0.5\baselineskip}{0.2\baselineskip}
\titleformat{\section}
   {\normalfont\larger[0.5]\sf\bfseries}{\thesection.}{1em}{}
\titleformat{\subsection}[runin]{\normalfont\sf\bfseries}{\thesubsection.}{1em}{}[~~]

\usepackage[margin=0.75in]{geometry}
\usepackage[T1]{fontenc}
\usepackage{graphicx}
\usepackage{float}
\usepackage{amsmath,amssymb,bm}
\usepackage{xcolor}
\usepackage{booktabs}
\usepackage{enumitem}
\usepackage{dcolumn}% Align table columns on decimal point
\usepackage{hyperref}% add hypertext capabilities
\usepackage{hycolor}
\hypersetup{colorlinks=true,citecolor=[rgb]{0,0.18,0.68}, 
urlcolor=[rgb]{0,0.18,0.68}, linkcolor=[rgb]{0,0.18,0.68}, final = true}
\usepackage[rightcaption]{sidecap}
\usepackage{physics}
\usepackage{mathtools}
\usepackage{relsize}
\usepackage{mathptmx}
\DeclareMathAlphabet{\mathcal}{OMS}{cmsy}{m}{n}
\usepackage[scaled=0.86]{helvet}

\usepackage{siunitx}
\DeclareSIUnit\angstrom{\text {Å}}

  \usepackage{natbib}
  \setcitestyle{super,citesep = {comma}}

  \def\be{\begin{equation*}}
  \def\ee{\end{equation*}}
  \def\ba{\begin{eqnarray}}
  \def\ea{\end{eqnarray}}
  
  \def\eref#1{(\ref{#1})}
  \def\eref#1{Eq.(\ref{#1})}
  \def\fref#1{Fig.~\ref{#1}}

  \def\bt{\textrm} %\textsf
  
  \def\nsb#1{\noindent\textbf{\bt{#1~}}}
  
  \definecolor{or}{RGB}{234,142,53}
  \definecolor{gr}{RGB}{150,150,150}
  \definecolor{bl}{RGB}{54,152,187}

  \def\rb{\mathbf{r}}
  \def\NN{\mathbf{D}}
  
  \def\dNi{S_{i}}

  \def\Keff{K_{\rm eff}}
  \def\AF{\mathrm{AF}}

  \def\AFave{\langle \AF \rangle}

  \def\CA{\mathrm{C}_{\alpha}}
  
  \def\dm{\delta_\mathrm{m}}
  \def\T{^\intercal}
  \def\Ht{\mathbf{H}}
  \def\Gt{\mathbf{G}}
  \def\Ft{\mathbf{F}}
  
  \def\gt{\mathbf{g}}
  \def\ft{\mathbf{f}}
  \def\ut{\mathbf{u}}
  \def\vt{\mathbf{v}}
  \def\rt{\mathbf{r}}

\let\epsilon=\varepsilon
\def\CC{{\mathcal{C}}}
\def\EE{{\mathcal{E}}}

\newcommand{\ech}{\CC}
\newcommand{\edf}{\EE}

  \newcommand{\ie}{\textit{i.e.}}
  \newcommand{\eg}{\textit{e.g.}}

  \definecolor{YKB}{rgb}{0.00,0.18,0.65}

\begin{document}

% \title{Internal protein motion: getting in the groove}
\title{\larger[+1]{\textsf{The Physical Logic of Protein Machines}}}

  \author{John M. McBride}
    \email{jmmcbride@protonmail.com}
    \affiliation{Center for Soft and Living Matter, Institute for Basic Science, Ulsan 44919, South Korea}
  \author{Tsvi Tlusty}
    \email{tsvitlusty@gmail.com}
    \affiliation{Center for Soft and Living Matter, Institute for Basic Science, Ulsan 44919, South Korea}
    \affiliation{Departments of Physics and Chemistry, Ulsan National Institute of Science and Technology, Ulsan 44919, South Korea}

\begin{abstract}
  Proteins are intricate molecular machines whose complexity arises from the heterogeneity of the amino acid building blocks and their dynamic network of many-body interactions. These nanomachines gain function when put in the context of a whole organism through interaction with other inhabitants of the biological realm. 
  And this functionality shapes their evolutionary histories through intertwined paths of selection and adaptation.
  Recent advances in machine learning have solved the decades-old problem of how protein sequence determines their structure. However, the ultimate question regarding the basic logic of protein machines remains open:
  How does the collective physics of proteins lead to their functionality? and how does a sequence encode the full range of dynamics and chemical interactions that facilitate function?
  Here, we explore these questions within a physical approach that treats proteins as mechano-chemical machines, which are adapted to function via concerted evolution of structure, motion, and chemical interactions.
\end{abstract}

\maketitle
\tableofcontents
\section{Introduction: protein machines}

  Proteins are often described as molecular machines,~\cite{kurzynskiProtein1997,albertsCell1998,kinbaraIntelligent2005,goodsellMachinery2009,flechsigSimple2019}
   but what does this mean?
  While a rigorous consensus definition is lacking, broadly speaking, machines are devices or objects that transform a physical system to perform a function. The operation of machines naturally involves some dynamics. Many machines have moving parts and exert time-varying forces.
  Other machines change their internal state as they perform computation, and often computation and motion are coupled, as in mechanical computers. 
  Proteins appear to fit into all of these criteria: they are complex objects
  which often have independently moving parts; motor proteins facilitate directional motion; binding to some molecules and not others is a form of computation; and they all evolve according to the task they perform -- their \textit{function}.

  In this article, we will treat proteins as deformable
  mechano-chemical machines, whose physical behavior and response to forces are
  encoded in their DNA sequences, and whose evolution is directed towards improved function. 
  We will address the question of why binding -- which underlies most protein
  functions -- requires \emph{motion}. We discuss how to quantify internal motion involved in function, and in particular, examine quantities derived from finite strain theory. We then turn to the mechanistic question of how specific
  functional motion is encoded in protein sequences.  
  Finally, we show how amino-acid substitutions lead to variations in protein structure that can be described as a kind of evolutionary motion captured by the same deformation metrics that describe the physical motion.
  Overall, the present study takes one step further toward understanding the physical logic of protein machines by proposing a general model of the genotype-to-phenotype map that takes into account the evolution of functional motion.

\begin{figure*}
\centering
\includegraphics[width=0.95\textwidth]{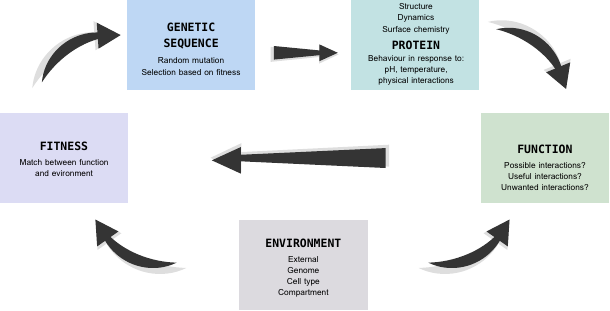}
\caption{\label{fig:fig1} 
  \textbf{The cycle of evolution of protein machines}.
  \textit{Genetic sequences} encode the protein's ensemble of molecular structures and its physical behavior.
  \textit{Protein behavior} is defined by structure, dynamics, and surface chemistry. These aspects of protein behavior vary depending on the
  physicochemical environment and in response to physical forces encountered by the proteins when interacting with other molecules.
  \textit{Functions} are constrained by the possible interactions a protein can have, but they are typically defined by the interactions that are \emph{useful} for the host organism.
  Proteins exist and function within a \textit{hierarchy of biological networks and environments}: proteins are often localized
  in specific cellular compartments and organelles, or only expressed in certain cell types; proteins often evolve to specifically interact with other molecules that are encoded by the host organism's
  genome, or encountered inside or outside the cell.
  The \textit{fitness} contribution of a gene depends on how well the function matches the environment and the needs of the host organism, and this, in turn, acts to bias the selection of random mutations in the genetic sequence.
}
\end{figure*}

\section{Protein biophysics and evolution}

  Understanding the logic of protein machines is an open challenge. A longstanding paradigm in biology has been that protein function is determined by structure, which is in turn determined by sequence. This has been a useful first approximation of the logic of proteins. But by now, we have a much more nuanced understanding, briefly summarized below, of the ongoing cycle of how genes encode proteins and their function, which in turn biases the evolution of genes (\fref{fig:fig1}).~\cite{tlusty2016self}

  A protein, of course, is not simply a static object captured by a single structure, but rather a dynamic one, whereby an ensemble of structures, their dynamics,~\cite{orellanaLargeScale2019,campitelliRole2020,namProtein2023,nussinovProtein2023} and surface chemistry intertwine to define a space of possible functions. A useful characterization of proteins must explain
  how these ensembles are shaped by the physico-chemical environment,~\cite{timasheffControl1993,schaeferPHDependence1997,meersmanProtein2006,obrienEffects2012,andersonAdaptive2021,pandeyPhysicochemical2022} and how
  they respond to physical forces when interacting with other molecules.~\cite{koshlandApplication1958,koshlandComparison1966,monodNature1965,savirConformational2007,mcbrideGeneral2022}
  Such responses can be characterized by a shift in the conformational landscape of a protein:~\cite{okazakiDynamic2008,cembranNMR2014,wanel22,stachowskiLargeScale2022}
  For example, conformations can change at binding sites due to induced fit,~\cite{koshlandApplication1958}
  or long-ranged, allosteric changes can occur throughout a protein;~\cite{monodNature1965}
  in extreme cases, binding can cause disorder-to-order transitions, whereby
  a disordered region can form a stable secondary structure upon binding.~\cite{gianniCoupled2016}

  The notion of function also needs to be re-examined, as some proteins can have many, varied functions (moonlighting),~\cite{jefferyProtein2018,bertoliniMultifacetedProtDB2023} and proteins can gain or lose function depending on the environment (gain of function, or pseudogenes).~\cite{zhuangMolecular2019,gerasimaviciusLossoffunction2022,harrisonStudying2002} If proteins are expressed in new cell types, or
  located in new sub-cellular locations,~\cite{hungProtein2011} they gain access to new sets of molecules to
  interact with.~\cite{noorenNEW2003,guptaMoonlighting2023} Varying external environments also provide a fluctuating landscape of
  interaction partners, which can switch proteins between functional and non-functional states (\eg, antibodies, anti-toxins).~\cite{marchalonisAntibody2006,hayesToxinsantitoxins2011} 
  Function is also defined not only by interactions
  with target molecules but, not less importantly, by interactions or lack thereof with off-target molecules. Proteins
  need to interact \emph{specifically} with certain molecules, while avoiding aggregation,~\cite{invernizziProtein2012}
  unwanted cross-talk in signalling,~\cite{vertCrosstalk2011} and enzymatic side-reactions.~\cite{vogeliNegative2018} Thus, to wholly understand protein function, one needs to understand the range of behaviors that a protein is capable of, know what molecules they will encounter, and how it all links with the fitness of the host organism. For this, we need a physical model of the full
  genotype-to-phenotype map, which takes into account the place of the protein in the biological circuitry and all relevant counterparts.

\section{Why do protein machines move?}

\subsection*{All proteins move by their physical nature.}
  The first theory of protein binding was the \textit{lock-and-key} model, which hypothesized rigid shape matching between
  enzymes and substrates.~\cite{fischerEinfluss1894} This model was proposed in 1894 by Emil Fischer, long before much was known about the size and complexity of proteins.~\cite{rastetterEnzyme1983}
  While geometry is indeed an important component of binding, by now, we are aware that proteins are large, flexible molecules that undergo internal motion (to which we refer hereafter simply as ``motion'' as we do not discuss whole-body translations and rotations).~\cite{buProteins2011}
  Proteins are bound to move simply because their structures are held together by relatively weak bonds, which can easily break and reform.
  Moreover, the linear-chain topology makes it impossible to optimize all
  of the bonds between amino acids.~\cite{ferreiroLocalizing2007,ferqr14} Thus the first answer to the question, "Why do proteins move?", is that they cannot help it. 
  However, beyond the physical inevitability, there is much evidence that motion in proteins is also beneficial and that the
  extent and direction of movement are selected through evolution and can be harnessed to perform functions.\\

\begin{figure}
\centering
\includegraphics[width=0.45\textwidth]{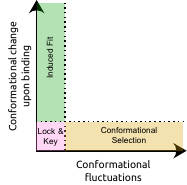}
\caption{\label{fig:fig2} 
  Mechanisms of binding described in a 2D diagram whose coordinates are the degree of conformational fluctuations in the absence of the substrate, and the magnitude of conformational change upon binding to the substrate.
  The classical models -- lock-and-key, induced fit, and conformational selection -- are shown as limiting cases.
  In reality, absolute rigidity and deformation-free binding are impossible -- proteins are inherently flexible and undergo thermal motion, while binding involves forces that will always induce some degree of conformational change.
}
\end{figure}

\subsection*{Mechanisms of binding involve movement.}
  The lock-and-key model was eventually updated to include movement, first through the induced fit mechanism, and later through conformational selection:~\cite{holyoakMolecular2013}
  In the induced fit model, proteins undergo conformational change only upon binding to substrate;~\cite{koshlandApplication1958}. In the conformational selection model, the unbound protein inherently samples many possible conformations,
   and binds when it achieves the optimally binding conformation.~\cite{straubRemarks1964,vertessyFluctuation2011,tsaiFolding1999}

  These two models can be described as extreme cases on a 2-dimensional continuum (\fref{fig:fig2}) where the two coordinates are 
  the degree of conformational flexibility in the absence of
  substrate (x axis) and the degree of conformational change upon binding (y axis).
  In reality, we expect that most binding events are described in the intermediate space in between the regions of induced fit, and conformational selection -- sometimes called the population shift model.~\cite{csermelyInduced2010,bucherInduced2011}
  Altogether these theories provide a comprehensive description of the mechanisms
  of binding in terms of protein motion, but the mechanisms have little to say about
  \textit{why} movement may be beneficial. In fact, if we purely considering molecular recognition,
  flexibility ought to be detrimental to binding, as more flexible proteins have greater
  reserves of conformational entropy that may be lost upon binding. To understand why
  proteins benefit from flexibility, we consider the fact that proteins
  exist in a milieu of many molecules, and they need to discriminate between
  target and off-target molecules. \\

\begin{figure*}
\centering
\includegraphics[width=0.95\textwidth]{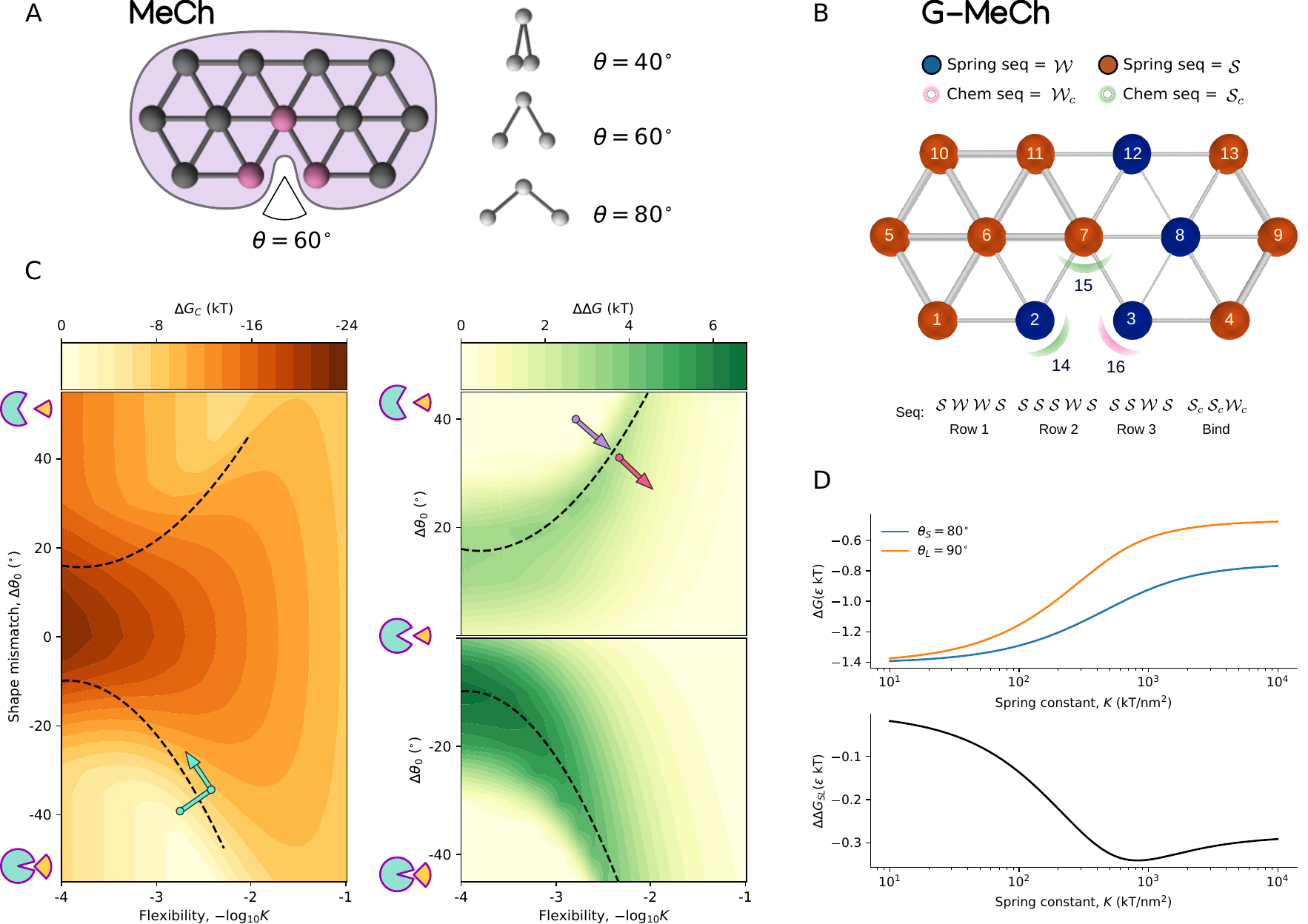}
\caption{\label{fig:fig3}
  A: Mechano-chemical discriminator model (\textbf{MeCh}):
  Coarse-grained model of a protein as spring network with a chemical binding site. In this example, the model includes $N=13$ nodes referred to as `amino acids' represent small sets of tightly bound coordinated amino acids.
  These amino acids interact with their neighbors via harmonic potentials.
  The binding site consists of three loci which bind to corresponding loci on wedge-shaped ligands.
  Spring constants are all equal to $K_{ij}=K$, and each binding loci can contribute a maximum of $\epsilon$ kT to the binding energy.\\
  B: Genetic, mechano-chemical \textbf{G-MeCh} discriminator model:
  The model is the same as \textbf{MeCh}, but now the energy and spring constants are a function of a genetic sequence $\gt =\{a_i\}$ with two amino acids species $\mathcal{S}$ and $\mathcal{W}$; springs can be either weak, intermediate, or strong depending on the amino acid identity of the neighbors, $K_{ij}(a_i,a_j)$; chemical binding energy can be either weak or strong depending on the sequence.\\
  C: Affinity and specificity phase diagrams in the \textbf{MeCh} model, as a function of flexibility and mismatch between the protein pocket ($\theta_0 = \SI{60}{\degree}$) and the ligand shape, $\Delta \theta_0$. Positive shape mismatch indicates that the binding pocket is larger than the ligand, and vice versa.\\
  D: Free energy of binding, $\Delta G$, for large ($\theta_L=\SI{90}{\degree}$) and small ($\theta_S=\SI{80}{\degree}$) ligands, and the difference in free energy $\Delta\Delta G_{SL} = \Delta G_{S} - \Delta G_{L}$, as a function of spring constant $K$ for the \textbf{MeCh} model.
}
\end{figure*}

\subsection*{Binding specificity necessitates internal motion.}
  The induced fit theory was motivated by observations on amino acid binding by tRNA synthetases which were incompatible with the lock-and-key theory. \citet{koshlandApplication1958}
  noted that differences van der Waals interaction energy are insufsdficient
  to explain how isoleucyl-tRNA synthetase selects its target, the amino acid isoleucine, over valine which differs only by a single methyl group (-CH\textsubscript{3}).
  The lock-and-key model can explain the selection of valine over isoleucine (by the corresponding valyl-tRNA synthetase) -- since isoleucine is larger, it can be sterically excluded. However, lock-and-key cannot explain the inverse selection since valine should fit into the binding pocket that also fits the larger isoleucine.
  The proposed solution was the induced fit model, whereby the binding of isoleucine must lead to a conformational change that unlocks the binding pose, while valine may not induce such a change. This theory, thus,
  offers one good reason why proteins might have evolved to move.

  The idea that movement is needed for binding specificity was revisited in
  the conformational proofreading theory.   ~\cite{savirConformational2007,SavirRecA2010,savirRibosome2013}
  This model gives a quantitative explanation of induced fit and conformational selection, which takes into account the inherent flexibility of proteins and the energetics of recognition.
  According to this theory, proteins may interact with off-target as well
  as target molecules due to conformational flexibility, if the two molecules are similar (by either induced fit or conformational selection). This leads to the prediction that, for molecular discrimination, the optimal protein may have some mismatch with the target: often, a small amount of mismatch between protein and target will slightly
  decrease the affinity to target, while disproportionally affecting binding to off-target ligands. \\
  
\section{Genetic-mechano-chemical model of protein machines}
  In our recent work, we produced a detailed quantitative model of protein binding that builds on these previous models, with the aim of elucidating the mechanistic determinants of binding specificity and how they are encoded in the evolution of the gene.~\cite{mcbrideGeneral2022}
  The model is presented in two variants -- a basic mechano-chemical machine (denoted \textbf{MeCh}), and a genetic-mechano-chemical machine (\textbf{G-MeCh}), which adds the genetic component. Here, we briefly describe the main concepts, while a full discussion of the model can be found in \citet{mcbrideGeneral2022}.
 
  We start with the simple variant of the model, \textbf{MeCh}, which takes into account flexibility, protein-ligand mismatch, and chemical binding energy (\fref{fig:fig3}A). For simplicity, we keep the initial structure constant (although see \citet{mcbrideGeneral2022} for a relaxation of this constraint), and model the protein as a 2-dimensional
  spring network of ``amino acids'', $a$, (each representing a tightly-connected bunch of amino acids) 
  on a hexagonal lattice. 

  Protein flexibility is described by connecting the nearest
  neighbors with harmonic springs, such that deformation incurs an energy penalty
  \begin{equation} \label{eq:edef}
    \edf = \frac{1}{2} \sum_{\langle i, j \rangle}  K_{ij} \left(r_{ij} - \ell_{ij} \right)^2~,
  \end{equation}
  where the summation $\langle i, j \rangle$ is over all pairs of amino acids
  $a_i$ and $a_j$ connected by a bond, $K_{ij}$ is the spring constant of the bond, $r_{ij}$ is its length, and $\ell_{ij}$ its equilibrium length.
  The initial configuration is not deformed, $\edf=0$, since all distances are set equal to the equilibrium bond lengths, $r_{ij} = \ell_{ij}$. 
  In the \textbf{MeCh} model, all spring constants are equal, $K_{ij} = K$.
  For simplicity, we treat ligands as rigid molecules. In reality, however, ligands can also be flexible, especially if the ligand is a protein or peptide, and the model can be extended to account for such effects.
  Flexibility also determines the conformational entropy of the protein, and conformational entropy loss upon binding is calculated under the assumption that binding results in a more rigid binding site;
  note that we also include a separate parameter to account for other sources of binding entropy change.

  Next, we consider protein-ligand mismatch and binding energy. In reality, mismatch and binding energy are inseparable, since amino acid substitutions at the binding site lead to concurrent
  changes in shape and chemistry. Nevertheless, we treat mismatch and binding energy as separable components, both for simplicity, and so that we may examine their separate contributions to binding specificity.
  Mismatch is entirely encoded in the shape -- the protein binding pocket has an angle of 
  $\SI{60}{\degree}$, and ligands are wedge-shaped with angles ranging from $\SIrange{5}{175}{\degree}$.
  Chemical binding energy is modeled as a short-ranged interaction between three amino acids at the binding pocket with three corresponding loci on the ligand,  
  \begin{equation}\label{eq:ech}
      \ech = \sum_{i} \epsilon_i \, 
      e^{-|\rb^b_i - \rb^p_i|^2/\sigma^2}~.  
  \end{equation}
  Here, $\epsilon_i$ is the energy scale of binding locus $i$ of the ligand, $\rb^p_i$ and $\rb^b_i$ are the positions, respectively, of the amino acids $a_i$ at the binding pocket and the ligand binding locus $i$, and we set the length scale of the interaction
  at $\sigma = \SI{0.3}{\nm}$. 
  In the \textbf{MeCh} model, the energy scale is the same for all binding sites, $\epsilon_i = \epsilon$.
  The energy scale, $\epsilon$, is equivalent to the maximum chemical binding energy afforded by a ligand's accessible chemical groups.

  While \textbf{MeCh} models a general, mechano-chemical discrimination machine, the \textbf{G-MeCh} model aims to describe more concretely proteins (\fref{fig:fig3}B). Therefore, instead of allowing the spring constants and energy scale to vary continuously, the mechano-chemical properties of the protein are defined by genetic sequence $\gt = \{a_i\}$ that determines the spring constants $K_{ij}(a_i,a_j)$ in \eref{eq:edef} and the binding interactions $\epsilon_i(a_i)$ in \eref{eq:ech}.  
  This results in a protein that undergoes discrete changes via mutations to its sequence, and is a heterogeneous rather than homogeneous spring network. 
  
  The simplicity of these models allows us to examine a large space of possible proteins, where we can calculate binding energy for a large range of ligands, and all possible genotypes. The complete survey of the genotype-phenotype map sheds light on the evolution of specific binding in proteins, and how this depends on the interplay between internal protein motion, shape mismatch, and binding energy.\\
  
\begin{figure*}
\centering
\includegraphics[width=0.95\textwidth]{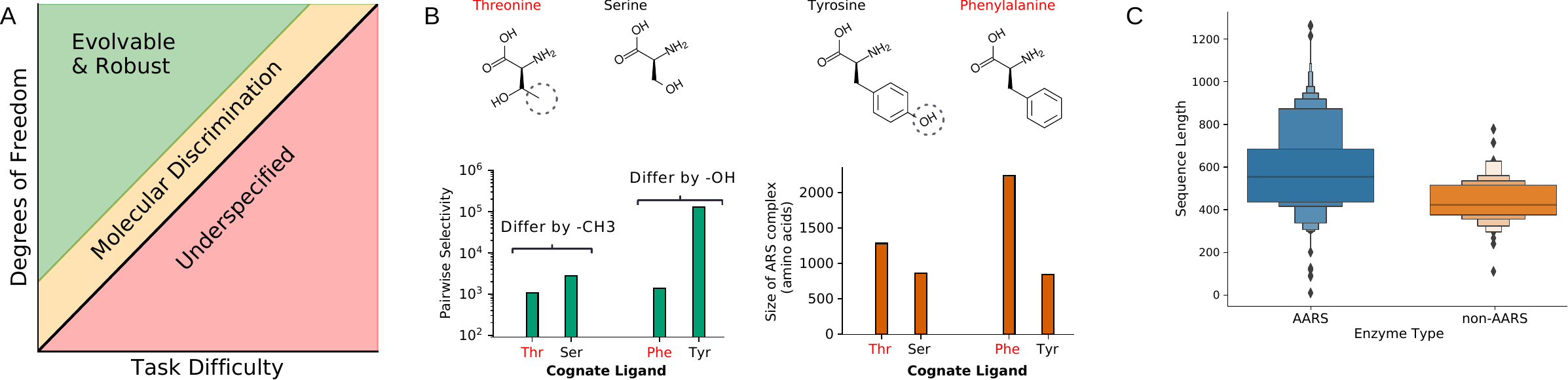}
\caption{\label{fig:fig4} 
  \textbf{Size-Specificity-Scaling Hypothesis}. A: The difficulty of a discrimination task imposes a constraint on the minimum sequence length
  that can lead to successful completion of the task. Sequences longer than the
  minimum length are more evolvable and robust.
  B: Amino-acyl tRNA-synthetase (AARS) pairwise selectivity (ratio of binding constant of cognate ligand to a similar non-cognate ligand) and number of residues in the AARS complex, for threonine, serine, phenylaline and tyrosine. Ligands that are more difficult to selectively bind to are highlighted in red.
  C: Sequence length distribution of AARS vs non-AARS proteins that also bind to amino acids.
}
\end{figure*}

\nsb{\textsf{Predictions from the genetic-mechano-chemical model:}} \\

\noindent (A) \textit{Suboptimality of Lock-and-key}:~
Lock-and-key binding is not the optimal strategy for evolving binding specificity;
  it only works when off-target molecules are larger (leading to steric exclusion), in the unrealistic case of extreme rigidity and perfect match between protein and ligand.
  Rigid proteins do not lose much conformational entropy upon binding, which can lead to very high binding affinity. Although affinity is higher for the matching ligand, similar off-target ligands will still bind even if the binding pose is not perfect (\fref{fig:fig3}C). This is the aforementioned insight that led to the development of induced-fit model.~\cite{koshlandApplication1958}\\

\noindent (B)~\textit{Mismatch is essential}:~ 
We show, in agreement with the \emph{conformational proofreading} model,~\cite{savirConformational2007} that some mismatch between protein and target is required for discrimination, in order to increase the energy penalty for binding to off-target ligands (\fref{fig:fig3}C).
  Through the \textbf{MeCh} model, we explain mechanistically that this is due to the non-linear nature of binding as a function of protein motion.
  Deformation energy (stretching of intramolecular bonds) increases at a greater rate for the off-target molecules than the target molecules. As a result, optimal specificity is achieved when the protein incurs a heavy deformation penalty which can be paid off by binding to the target molecule, but not to the off-target molecule (\fref{fig:fig3}D).\\

\noindent (C)~\textit{Discrimination requires finetuning of multiple coupled factors:}  In principle, better discrimination is achieved by the most rigid proteins, but the benefit of high rigidity is marginal. Instead of maximizing rigidity, it is much more important for proteins to carefully tune the interplay between (i) ligand mismatch, (ii) the energy scale of binding (\ie, number, and strength of potential bonds), and (iii) flexibility (\fref{fig:fig3}C). As a result, there are no monotonic relations between specificity and any of these three components. 
  
  This novel prediction is at odds with conventional wisdom in
  the fields of enzyme and antibody evolution, where it is often assumed that specificity monotonically decreases with increasing flexibility.~\cite{zimmermannAntibody2006,pabco18,petrovicConformational2018,vanregenmortelSpecificity2014,mishraInsights2018} This prediction can explain reportedly `counterintuitive' behavior in a family of esterases, where specificity increases with increasing flexibility.~\cite{nutschelPromiscuous2021}
  Since the esterases in question have large, open binding pockets, and specificity is found to increase with decreasing binding pocket size,~\cite{martinez-martinezDeterminants2018} we can infer the approximate location of these esterases on the phase diagram (\fref{fig:fig3}C, purple arrow), and predict that further increases in flexibility will eventually lead to a decrease in specificity (\fref{fig:fig3}C, red arrow).
  
  Likewise, the predictions of the affinity-specificity phase diagram are compatible with the results of a study
  on antibody evolution, where an evolutionary trajectory selecting for affinity first led to increased
  flexibility, followed by decreased flexibility (\fref{fig:fig3}C, cyan arrows).~\cite{ovcel18}
  Our model predicts that this evolutionary trajectory is expected when the initial antibody is a poor match for the substrate.
  In general, we speculate that the reason for the inferred monotonic relation between flexibility and specificity could be a bias from observing wild-type proteins; if proteins are optimized for specificity, then they will fall along the optimal line (\fref{fig:fig3}C, dashed line), along which flexibility and specificity are proportional to each other.\\

\noindent (D)~\textit{Distant residues are essential}:~
A key finding from the \textbf{G-MeCh} model about the role of flexibility in molecular discrimination is that residues \emph{far} from the binding site are important for discrimination.
  While mutations at distal residues have relatively small effects on binding modes, compared to residues at the binding site, these small effects are exactly what is needed for achieving precision in binding interactions. \\

 \noindent (E)~\textit{Why are proteins so large? }~ 
 The \textbf{G-MeCh} model shows that 
  \emph{larger} proteins are able to discriminate between a greater
  range of ligands, and they are also more evolvable and robust.~\cite{mcbrideGeneral2022}
  Proteins are considered \emph{evolvable} if they can gain a new phenotype (\eg, an enzyme that is repurposed to catalyze a new reaction through a few mutations). Proteins are \emph{robust} if they can withstand many mutations while maintaining functionality.
  The robustness and evolvability of larger proteins can be explained by the findings that (i) good specificity is achieved by precise tuning of flexibility, mismatch, and chemical binding energy, and that (ii) greater precision is achieved through fine-tuning via mutations at residues distal to the binding site. Extra amino acids in the protein sequence afford greater degrees of freedom to tune for accurate discrimination. 
  
  This finding leads to the hypothesis that the difficulty
  of achieving a certain degree of specificity between ligands sets a 
  minimum size constraint on proteins. In other words, the minimum size of a functional protein
  scales with the specificity requirements of the discrimination task.
  We call this the \textit{size-specificity-scaling hypothesis}. 
  Proteins need to be at least as big as some minimum size,
  and being larger increases evolvability and robustness, thus facilitating
  evolution (\fref{fig:fig4}A). The requirements of specificity thus can explain why proteins
  are so large -- they are not simply structural supports for active sites, but
  rather expansive molecular machines that are precisely calibrated for function.

  The size-specificity-scaling hypothesis might be hard to test
  because the task difficulty lacks a rigorous mathematical definition.
  But we can circumvent this issue by breaking down task difficulty into two components:
  the difficulty of specifically binding to one molecule over another, $X_{\rm diff}$, and the required specificity, $\Delta\Delta G$, between the two molecules.
  While $X_{\rm diff}$ is hard to define, we can test the hypothesis
  even without a definition by controlling for $X_{\rm diff}$.
  
  Thankfully, there exists a natural test case for this hypothesis -- amino-acyl tRNA synthetases (AARS).
  For certain amino acids, we can perform pairwise comparisons where we know that $X_{\rm diff}$ is higher in one case than the opposite case. For example, distinguishing tyrosine from phenylalanine is easier
  than the reverse task due to the increased chemical binding energy thanks to
  the extra hydroxyl group on tyrosine. Likewise, distinguishing threonine from serine is easier than the reverse case because the extra methyl group on threonine does not afford much greater binding energy, but it allows serine to be favoured through steric repulsion of threonine.
  
  As predicted by theory, the AARSs with easier tasks not only have higher selectivity, but are also smaller (\fref{fig:fig4}B).
  Furthermore, non-AARS amino-acid binding proteins are much smaller
  than AARSs (\fref{fig:fig4}C), which is what we expect given that AARSs
  require much greater selectivity, due to prohibitively high cost of incorrect protein translation.
  We will return to this in the discussion section with a conceptually simple,
  and feasible experiment to test this hypothesis.\\
  
\subsection*{Recap: why do proteins move?}
  Molecular discrimination -- which is more difficult than mere recognition --
  is predicted to benefit from internal motion in proteins by the theories
  of induced fit, conformational proofreading, and our genetic-mechano-chemical model of proteins~\cite{koshlandApplication1958,savirConformational2007,mcbrideGeneral2022}, which provides a physical basis to this hypothesis.
  Many essential protein functions require carefully calibrated functional motion:
  AARSs need to discriminate between amino acids with exceptional precision, and transcription factors
  need to discriminate between a huge range of similar targets.~\cite{jolmaDNABinding2013} Antibodies
  need to bind to pathogens while avoiding products of the host organism,
  otherwise leading to auto-immune disorders,~\cite{mahlerEpitope2010} and enzymes need to simultaneously
  optimize for both binding to substrates and also populating the transition state,
  while simultaneously facilitating product release.~\cite{peracchiLimits2018,rivoireGeometry2020,rivoireHow2023}

  While we so far presented evidence for how specificity requirements necessitate functional motion in proteins,
  there may be other important reasons for the evolution of conformational dynamics,
  such as promiscuity,~\cite{copleyShining2017} signalling,~\cite{wodakAllostery2019} and evolvability.~\cite{pabco18,johanssonStructural2018,campitelliRole2020}
  Specificity is essential to many protein functions, but there are also many promiscuous ``generalist'' proteins with multiple interaction partners.~\cite{nobeliProtein2009,cumberworthPromiscuity2013}
  For example, proteins involved in host defense, such as broadly-neutralizing antibodies,~\cite{cortiBroadly2013,chenBroadly2023}
  or promiscuity in antibiotic resistance,~\cite{frohlichEvolution2021} benefit from the ability to ward off diverse attacks.
  Signalling is aided by long-ranged, allosteric dynamics, which enable transduction of force across membranes,~\cite{grovesMolecular2010} and switch-like inhibition/activation.~\cite{kernRole2003}
  Furthermore, it has been hypothesized that promiscuity is an evolutionary driver for functional diversification.~\cite{khersonskyEnzyme2006,campbellRole2016}
  To summarize, proteins move first because they have to, but the precise way in which they move is dictated by evolutionary history and biological function.

\begin{figure*}
\centering
\includegraphics[width=0.95\textwidth]{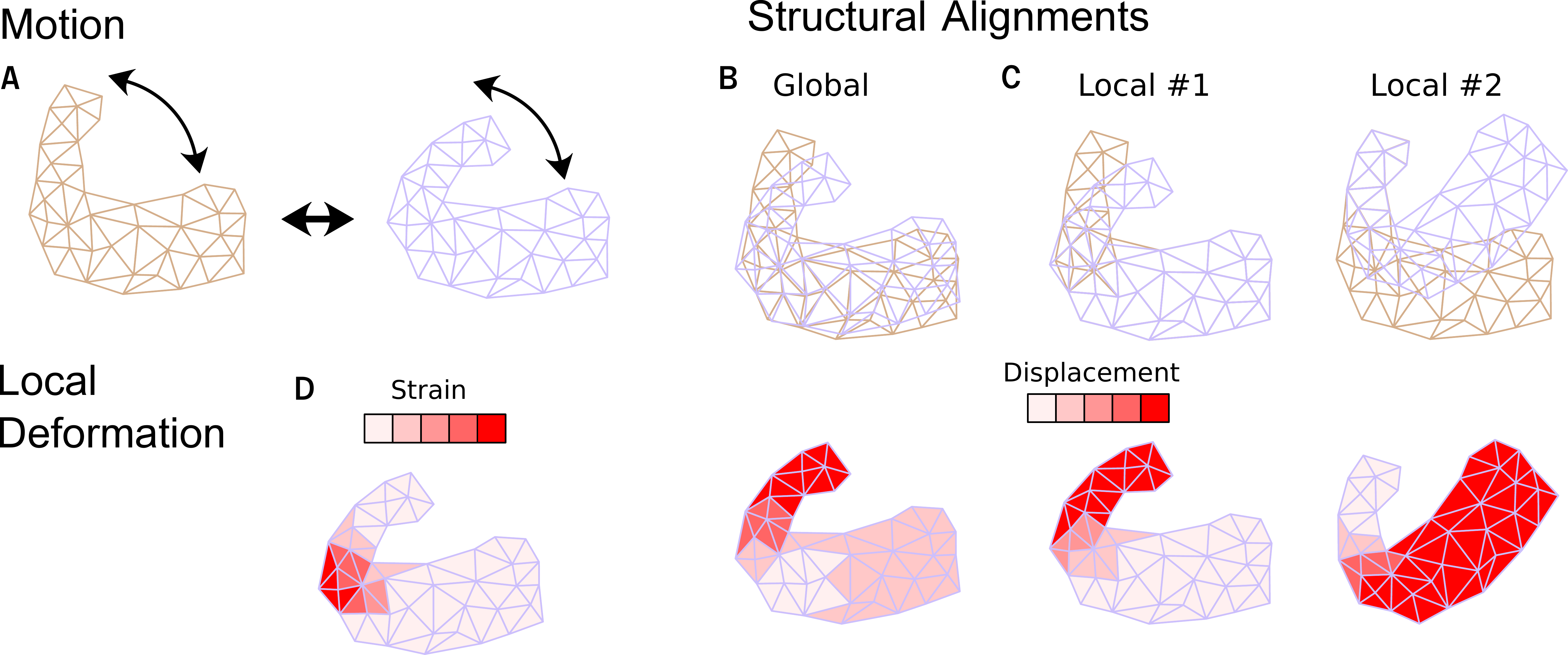}
\caption{\label{fig:fig5} 
  \textbf{Quantifying protein motion and deformation.}
  A: The internal motion of protein includes separate movements of its sub-units. 
  B,C show possible alignments of the original and deformed structure, where the top figures are the alignment and the bottom ones show the corresponding displacement. B: \emph{Global} measures of motion capture large-scale motion relative to some reference configuration, but they misrepresent local motion due to the use of global alignments. 
  C: \emph{Local} alignments of only a subset of amino acids illustrate this. B, and C demonstrate that measures derived from displacement, such as RMSD, strongly depend on the alignment. 
  D: An alternative approach is to measure \emph{strain}, which is a local measure of deformation. Strain is low if parts of a protein move together as a rigid body, and high if their relative positions change, which implies stretching, compressing and breaking of bonds, and the formation of new bonds. Thus, strain naturally captures local deformation relevant to function. 
  }
\end{figure*}

\section{Quantifying internal motion}
  Protein motion can be measured using various techniques that access specific timescales, spatial resolution, and different degrees of molecular and atomic resolution. No one method is superlative, as they each have strengths and deficiencies.
  Here, we are interested in the methods that provide structural or dynamical information in terms of the full atomic coordinates of proteins, both experimentally (X-ray crystallography, NMR spectroscopy, cryo-EM)
  and computationally (molecular dynamics, elastic network models, structure prediction algorithms). In particular, we will discuss how one should mathematically approach measuring differences between structures.\\

\subsection*{What is a good metric of internal motion?}
  When measuring motion between protein conformations, we need to specify both the set of atoms to compare, and the frame of reference in which to compare them.
  We can consider \emph{local} motion between neighboring amino acids, or the \emph{global} motion that compares all amino acids. 
  Most methods involve some form of \emph{structural alignment}, which defines a frame of reference. The alignments are needed to eliminate differences in atomic coordinates due to whole-body translations and rotations. However, there is no single, universally correct method of alignment, since proteins consist of many moving parts that can individually undergo translations and rotations (\fref{fig:fig5}).
  Ideally, one desires methods for measuring protein motion on both local and global scales, that are insensitive to the choice of alignment.

  The most common metric of protein structural change is root mean square deviation (RMSD), which measures the average displacement between atoms in a target and reference structure, after aligning the target to the reference structure; this is a useful measure of global motion.~\cite{kuf12}
  We note that in order to get residue-level information on global motion, one can calculate correlation functions or covariance. These have been shown to successfully predict long-range correlations that do not depend on the structural alignment.~\cite{tangCritical2017,tangLongrange2020}

  For a local measure of motion one can examine the displacement per amino acid. However, the main issue with this measurement is that displacement can be large in areas that do not undergo any local rearrangements (\fref{fig:fig5}), so certain residues exhibit large shifts from the reference structure, while locally retaining exactly the same configuration.
  
  Many methods have been developed to overcome this problem, yet they all share a common weakness. Methods such as template-modelling score, global distance test, and local distance difference test, are designed to measure similarity instead of difference.~\cite{zemna03,zhapr04,marbi13} As a consequence, they underweight outliers (flexible regions tend to move a lot, and this is to be ignored) and are normalized to produce a score instead of a true \emph{metric} in the mathematical sense.
  In the following, we will consider how to measure local motion in proteins in a way that avoids the inherent problems associated with internal motion, while also maintaining consistency across different magnitudes of motion.

\subsection*{Finite strain theory for proteins.}
  For a more rigorous metric of local motion, we look to finite strain theory, which gives us a natural framework for analyzing deformation (\fref{fig:fig5}).~\cite{lublinerPlasticity2008}
  Strain can be broken down into affine and non-affine strains. The affine component includes compression/expansion, shear, and twist motions, which can be described as linear transformations of the coordinates,  and the non-affine component includes all the non-linear residuals. 
  The affine part of strain is captured by the deformation gradient tensor, $\Ft$, which describes the transformation between the two configurations, $\rt$ and $\rt'$, as
  \begin{equation}\label{eq:strain_tensor}
      \rb' = \Ft\,\rb~,
  \end{equation}
  The diagonal elements of $\Ft$ correspond to expansion or compression, while the off-diagonal elements relate to shear and twist motions. 
  
  In cases where deformation is a smooth function in space (\eg, continuum mechanics), deformation is entirely affine, and the linear relation \eqref{eq:strain_tensor} holds.
  In discrete systems, as is found at atomic nanoscale, this is no longer the case, and the residual of \eqref{eq:strain_tensor} 
  (\ie, the deviation from the linear relation) reveals a non-affine component of the overall strain.~\cite{broederszMolecular2011,wangAffine2017,prakashchandNonaffine2020}
  Here, we explain how to calculate the affine, non-affine quantities for proteins.~\cite{mitchellStrain2016,eckmannColloquium2019,mcbrideAlphaFold22022}

  We consider the case of measuring strain between two protein conformations, which are specified by the positions of their $\CA$ atoms, $\rt$ and $\rt'$ (each amino acid has one such atom). 
  Strain is a local quantity, which we measure at each amino acid $i$, by comparing the relative positions of the neighbors of $i$, $j \in N_i$, between conformations $\rt$ and $\rt'$.
  Neighbors of residue $i$ are those residues that are within a distance $\gamma$ of residue $i$; \ie, if
  $r_{ij} = |\rb_{ij}| = |\rb_i - \rb_j| \leq \gamma$.
  We then define a neighborhood tensor $\NN_i$ as the tensor of the $n_i=|N_i|$ distance vectors $\rt_{ij}$. Thus, we can compute the deformation gradient tensor at each amino acid $i$ of a protein, $\Ft_i$ as the solution of linear relation,
  \begin{equation}\label{eq:strain_tensor_protein}
      \NN_i' = \Ft_i \, \NN_i ~.
  \end{equation}
  We find $\Ft_i$ using linear regression, which minimizes the deviation from \eref{eq:strain_tensor_protein} due to non-affine deformation.  Then, we can extract the bulk (diagonal), shear (off-diagonal), and non-affine (the residual deviation) components.
  For proteins, it appears that the magnitude of the bulk component is generally lower than the magnitude of the shear component~\cite{mitchellStrain2016}, which is in turn lower than (yet, the same order of magnitude and correlated with) the non-affine component.~\cite{mcbrideAlphaFold22022}

  To formulate a single metric for measuring deformation that incorporates both affine and non-affine components of strain, we use what we term \textit{effective strain} (ES). This is simply the average relative change in distance vectors between neighbors,
  \begin{equation}
      \dNi = \left\langle \frac{\abs{ \Delta \rb_{ij} }}{r_{ij}} \right\rangle = \frac{1}{n_i}\sum\limits_{j \in N_i}
      \frac{\abs{ \rb_{ij} - \rb_{ij}' }}{r_{ij}}~. 
  \end{equation}
  At each amino acid, calculation of ES first requires a local structural alignment of the neighborhood tensors of the deformed and reference states; the corresponding rotation is found using the Kabsch algorithm.~\cite{kabac76}
  % ES is similar to the frame-aligned point error (FAPE) used to train AlphaFold2,~\cite{jumna21} where the main difference is that the contribution to the ES from each neighbor is normalized by the distance $r_{ij}$, while FAPE is in units of length.
  % A faster-to-calculate measure that avoids the alignment step is the local distance difference (LDD); although LDD is not sensitive to rotations of neighbors relative to each other, it has been found to be highly correlated with ES.
  See \citet{eckmannColloquium2019} and \citet{mcbrideAlphaFold22022} for more details on calculating strain.
  All of the above methods have been implemented in the python package \textit{PDAnalysis} (\href{https://github.com/mirabdi/PDAnalysis}{github.com/mirabdi/PDAnalysis}).

\begin{figure*}
\centering
\includegraphics[width=0.95\textwidth]{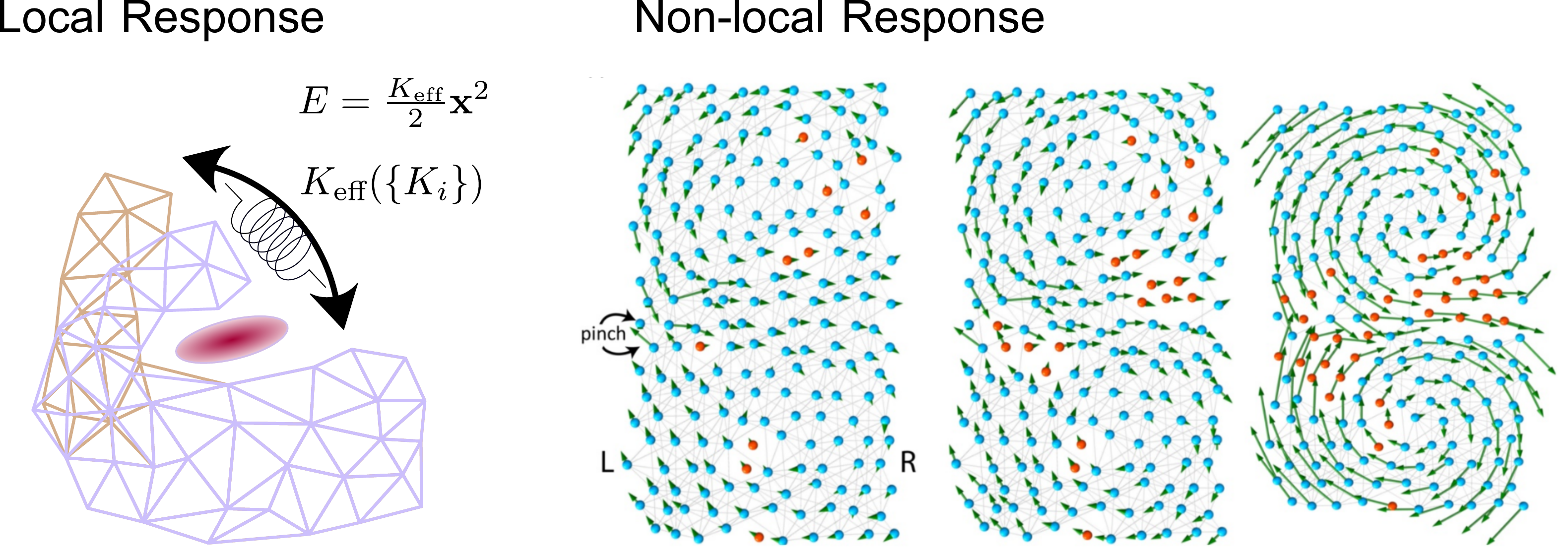}
\caption{\label{fig:fig6} 
  \textbf{How is movement encoded in protein structure?}\\
  \textit{Local Response:} We consider a protein as an elastic network where neighbors interact via harmonic potentials. A local response occurs when ligand binding causes displacement of amino acids at the binding site. This local deformation is apportioned non-locally throughout the spring network, thereby conveying long-range elastic forces.
  For any specific displacement vector $\mathbf{x}$, the spring network can be described as a single spring with an effective spring constant $\Keff(\{K_i\})$, where $K_i$ are the individual spring constants of each spring $i$.
  This means that mutations far from the binding site can affect the response of the protein to binding by altering $\Keff$ for $\mathbf{x}$.
  \textit{Non-local Response:} Protein responses to ligand binding can be long-ranged (\eg, allosteric regulation, signal transduction), such that a molecule binds at one site and causes motion at a 
  distal site. In a 2-dimensional physical model of proteins,~\cite{duttaGreen2018} with two types of amino acids (HP model: 'H', blue, forms strong bonds; 'P', red, forms weak bonds),
  the protein is evolved so that a pinch force $\ft$ at one end produces a response $\vt$ at the other end. The typical solution obtained through
  evolution is a channel of weakly bonded amino acids that undergo high strain in response to a pinch, which allows force to be transduced through the protein.\\
}
\end{figure*}

\section{Physics and evolution of protein motions}

  Proteins are easily deformable due to the soft nature of the non-covalent
  bonds holding them together, which are easily stretched or broken.
  Breaking of multiple bonds can occur cooperatively which leads to bistability,~\cite{ferrellSelfperpetuating2002}
  or even multiple conformational states separated by energy barriers.~\cite{wangExploration2012}
  Bonds stretch and break due to physical forces
  from binding and colliding with other molecules, or from thermal motion in the solvent.  
  Here, we will consider protein deformation as a response to local forces (\eg, induced fit), and as a response to non-local forces (\eg, allosteric inhibition/activation). 
  We will examine two models,~\cite{tlustyPhysical2017,duttaGreen2018,mcbrideGeneral2022}, where, for simplicity, only elastic forces are considered without explicit bond breaking.\\

\subsection*{Elastic network models.}
  Motion at binding sites in response to force, whether large-scale or minute, is determined by the many-body interaction network of the amino acids, which is encoded in the gene.
  The amino acids closest to the binding site are most critical
  for protein function as they largely determine the geometry and chemistry of the binding site. Nevertheless, mutations at distal residues have also been shown to affect binding and catalysis, independently from confounding effects on folding stability.~\cite{marsc21}
  In some cases, these residues have been found to impact dynamics by controlling the probability of conformational sub-states.~\cite{tousignantProtein2004,naganathanModulation2019}
  
  One possible mechanism for such non-local effects of distant amino acids can be understood by treating a protein as an elastic network, in which mechanical forces are transmitted over long ranges.~\cite{tirionLarge1996,baharDirect1997}
  To intuit the basic idea, consider the very simple case of a 1D protein made of a linear chain of $N$ springs with constants $\{K_i\}$. Assume that successful binding requires that amino acids at the binding site are displaced by $x$ relative to its neighbors.
  If this deformation remains localized in a single bond with spring constant $K_1$, then the elastic energy is $\edf_1 = K_1 x^2/2$. But if we let this displacement distribute over the $N$ springs, the overall energy relaxes to $\edf_N = \Keff x^2/2$, where the effective spring constant is  the harmonic mean, $\Keff(\{K_i\}) = [\sum_{i=1}^{N}{K_i^{-1}}]^{-1}$, and the motion of each spring is inversely related to its strength, $x_i \sim 1/K_i$. 
  Proteins are much more complex, of course, as they consist of networks of interconnected bonds, also in parallel, but the basic principle remains the same: the response to any localized force can be described using a single effective spring constant, which is a simple function of all spring constants and network connectivity, $\Keff(\{K_i\}) (\fref{fig:fig5})$.~\cite{eyalMolecular2008,mcbrideGeneral2022}
  This explains how the internal motion $\mathbf{x}$ is encoded in the topology and strength of the bonds. Next, we explain how this can be encoded in a protein's gene sequence.
  
  In our \textbf{G-MeCh} model, we use a binary sequence of amino acids whose identities encode the strengths of springs connecting the $n \leq 6$ adjacent amino acids.~\cite{mcbrideGeneral2022}
  This means that each mutation can alter the effective strength $\Keff$ by a discrete amount. Since each amino acid affects $\Keff$ to a different degree, larger proteins with more amino acids offer more ways to fine-tune $\Keff$, which enables more precise interactions between proteins and ligands. In summary, the simple spring network model explains (i) why and how distant residues are involved in deformation due to binding, (ii) how
  the collective properties of many amino acids can be reduced to
  a low-dimensional representation in terms of functional motion, and (iii) how this is encoded in sequence.~\cite{tlusty2016self, duttaGreen2018,eckmannColloquium2019,eckmannDimensional2021} \\

\subsection*{Protein sequences encode non-local response.}
  Since forces can propagate throughout proteins, they could evolve functions, such as signal transduction and
  regulation, involveing long-ranged communication between amino acids, known as allostery.~\cite{monodNature1965,koshlandComparison1966}
  Much of our understanding of allostery comes from elastic network models and molecular dynamics (MD) simulations, where it has been shown that the way force propagates along a network depends on its topology,~\cite{yanArchitecture2017,rocksDesigning2017,flechsigDesign2017,wodakAllostery2019,thirumalaiSymmetry2019}
  and that allosteric transduction pathways in proteins are facilitated by low-energy, slow modes in the intrinsic dynamics of proteins.~\cite{zhangIntrinsic2020}
  Here, we will revisit this question with a focus on how proteins \emph{evolve} such allosteric function, through a minimal model of protein function and evolution.~\cite{tlustyPhysical2017,duttaGreen2018}

  Models are always some compromise between “realism” (how close they are to real proteins) and simplicity, which allows fast computations and the ability to see something in the high-dimensional spaces of protein evolution. The main difficulty in studying protein evolution is the ``curse of dimensionality'': the need to
  obtain large enough statistics of long evolutionary paths, such that we can filter out the noise of random fluctuations.
  Nevertheless, if we wish to understand the physical response
  of the system to localized forces (i.e. binding induced allostery),
  there exists an old physical concept that achieves exactly this: the Green function.~\cite{greenEssay1828}
  
  The Green function describes the motion of the whole system as a response to a local perturbation: If a ligand binds at a certain site, many amino acids will move, depending on the network of intramolecular forces, and the interactions with the salty water at the protein surface (\eg, the formation of hydrogen bonds with water molecules).  
  As described below, we use Green's functions to build a simplified model of protein as an evolving elastic network whose interactions are encoded in a gene (\fref{fig:fig6}).~\cite{duttaGreen2018,eckmannColloquium2019}
  To make the model tractable for evolutionary simulations, we make the following simplifications:
  (i) The protein is described as a 2-dimensional network. (ii) There are only two types of amino acids: H - hydrophobic and P - polar, following the HP model.~\cite{dillTheory1985} The genetic space contains genes $\gt$, which are strings of zeros and ones.
  (iii) Intra-protein interactions are harmonic, and bonds are at minimum energy in the ground state.
  All three assumptions can be relaxed if one wishes to
be more realistic about a certain aspect.

  To formally define the Green function, we consider a toy protein made of a harmonic spring network. The elastic response of the protein is captured by an elasticity tensor $\Ht$. Basically, this tensor is a generalized spring constant $\Ht$ that tells us what force field $\ft$ occurs in response to a motion field $\ut$, through
  the linear relation, $\ft  =  \Ht \, \ut$ (the fields $\ut$ and $\ft$ assign force and displacement vectors to each amino acid of the protein). The elasticity tensor $\Ht$ depends on the strength and $\{K_i\}$ connectivity of the spring between the amino acids, which are encoded by the gene.  
  Therefore, we can write $\Ht$ as a function of the gene $\Ht(\gt)$.
  From $\Ht$, we can derive the effective spring constant for any displacement $\Keff(\{K_i\})$.
  As usual, the Green function is then the inverse tensor $\Gt(\gt)$ = $\Ht(\gt)^{-1}$, which is evidently also a function of the gene. The function $\Gt(\gt)$ solves the inverse problem: what is the motion
  induced by a force field $\ut(\gt)$? \ie, $\ut(\gt) = \Gt(\gt) \, \ft$.~\cite{duttaGreen2018,eckmannColloquium2019} 

  In this theoretical picture, the performance of the protein is measured by the similarity of the induced response $\ut(\gt) = \Gt(\gt)\,\ft$ and a reference functional collective mode $\vt$, which can represent for example the motion during induced-fit binding. 
  We can therefore define a performance measure $F$ (kind of “fitness”) of our toy protein as the overlap, $F(\gt) = \vt\T \cdot\ut = \vt\T \Gt(\gt)\,\ft$. Thus, in this simple model, we have an explicit \emph{genotype to phenotype map}, $\gt \to \ut(\gt) = \Gt(\gt)\,\ft$, and 
  a corresponding \emph{fitness landscape}, $\gt \to F(\gt) = \vt\T \Gt(\gt)\,\ft$.

  With this simple mathematical description, we can perform many simulations
  of evolutionary trajectories, starting from a random gene $\gt_0$ until we find a functional gene $\gt^*$, where functionality means having a fitness above a certain threshold, $F(\gt^{\ast}) \ge F^{\ast}$. A typical simulation of how evolution reaches a functional protein is shown in (\fref{fig:fig6}).  
  In general, the model explains how functional proteins
  with regions of localized high strain (a.k.a. ``shear bands'') emerge during evolution. In this physical picture, mutations
  induce local perturbations in the amino acid network, which shape the genotype-to-phenotype map.
  In particular, mutations induce significant structural deformation at high-strain regions (\fref{fig:fig6}),
  and the evolutionary interaction between these mutations (epistasis) corresponds to physical interaction among these ``defects''. This epistasis shapes the genotype-to-phenotype map.

\section{Protein evolution as effective internal motion}

  Like \emph{physical} perturbations (binding, physicochemical environment changes), \emph{evolutionary} perturbations -- that is, structural changes resulting from mutations -- can induce effective motion in the conformational free energy landscape of a protein, albeit on vastly different timescales.
  Evolutionary perturbations may cause allosteric changes that affect structure rather than (or in addition to) dynamics, and these perturbations may have a physical basis similar to that of perturbations induced by physical forces.~\cite{echcp08,tlustyPhysical2017, tangDynamicsEvolution2021}
  Despite mounting evidence, the exact mapping between the response to evolutionary perturbations and physical perturbations is still unclear. The mapping is easy to deduce in simple models but is much more elaborate in real proteins. In the following, we consider the effect of evolutionary perturbations on structure, using effective strain (ES) as a natural metric.

\begin{figure*}
\centering
\includegraphics[width=0.95\textwidth]{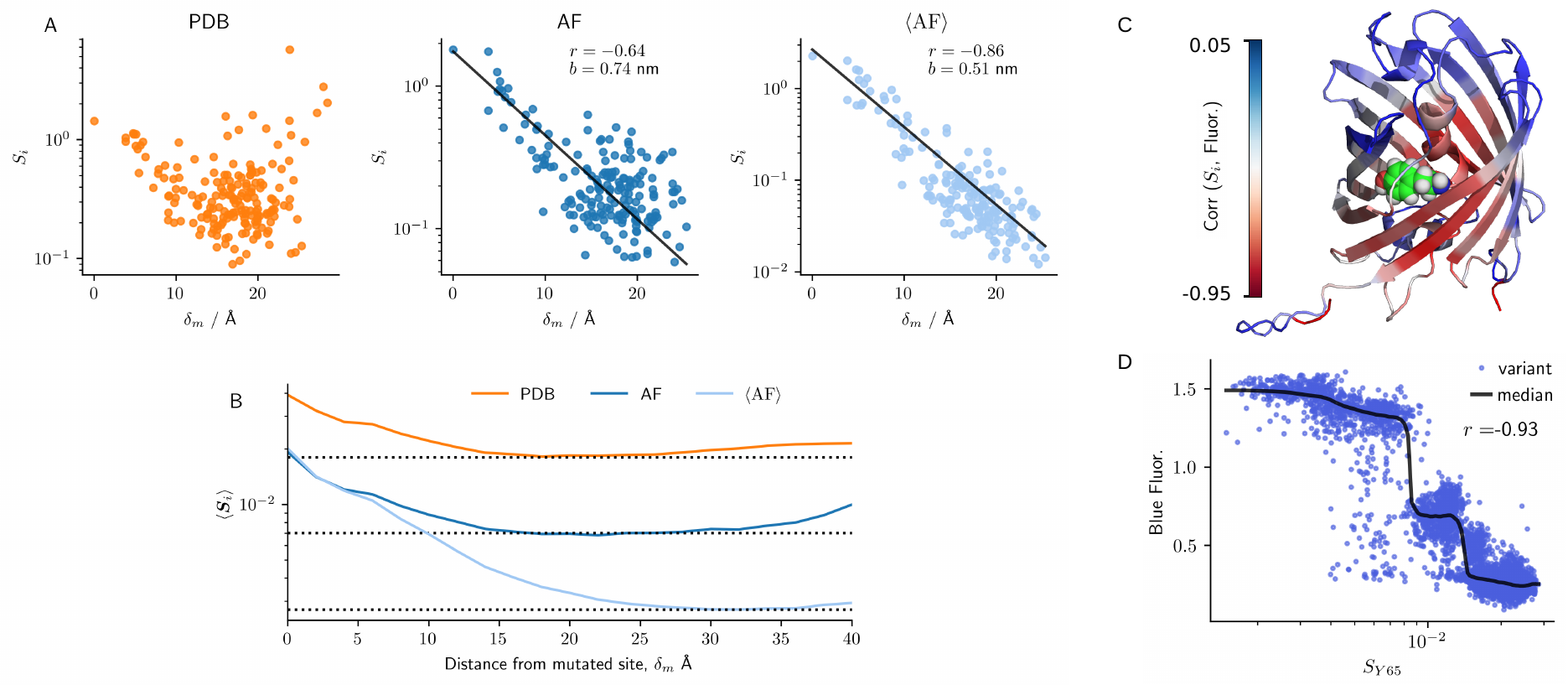}
\caption{\label{fig:fig7} 
  \textbf{Protein structure evolution as effective motion}.
  A: Effective strain $\dNi$ vs distance from a mutated site $\dm$, calculated using wild-type (WT) CypA (\texttt{6U5C\_A}) and a double mutant (S99T, C115S, \texttt{6BTA\_A});
  results are shown for PDB structures, AF-predicted structures, and effective strain (ES) calculated using ``averaged'' structures obtained from multiple AF predictions ($\AFave$);~\cite{mcbrideAlphaFold22022}
  black lines indicate fits to the function $\dNi = ae^{-\dm/b}$ .
  B: Average strain $\langle \dNi \rangle$ (over \num{12627} pairs of structures~\cite{mcbrideAlphaFold22022})
  vs $\dm$ for PDB, AF, and $\AFave$ structures.
  C: Blue fluorescent protein (BFP) colored by how well deformation at each residue $\dNi$ correlates with fluorescence; chromophore-binding site (Y65) is shown as spheres.
  D: Fluorescence vs ES (compared to WT) at residue Y65 for \num{8191} BFP variants; moving median line and Pearson's $r$ are shown.
}
\end{figure*}

  To this end, we use DeepMind's protein structure
  prediction algorithm, AlphaFold (AF), to predict the effect of mutation on structure as a strain field.\cite{mcbrideAlphaFold22022}
  As a case study, we measure strain between the  WT CypA and a double mutant (S99T, C115S),
  using three types of structural data (\fref{fig:fig7}A): structures from the protein data bank (PDB); AF-predicted structures;
  an ``average'' structure ($\AFave$) created by averaging over local neighborhoods of multiple AF-predicted structures,
  which has been shown to give a more precise estimate of deformation due to mutation (as opposed to stochastic noise).\cite{mcbrideAlphaFold22022,mcbrideAlphaFoldpredicted2023}
  We see that the strain in PDB structures depends on distance from the nearest mutated residue, $\dm$, within the first \SI{1}{nm}; beyond this high strain is an artefact of high variance between
  repeat measurements in flexible regions of PDB structures.~\cite{mcbrideAlphaFold22022}
  In contrast, AF has much lower measurement variance,~\cite{mcbrideAlphaFold22022}, and we see a clear exponential dependence of strain on $\dm$; this dependence is exceptionally clear for strain calculated using $\AFave$ structures.

  To understand the range of mutation effects, we report analyses of
  a set of \num{3901} proteins taken from \citet{mcbrideAlphaFold22022}, in which pairs of proteins differ by up to three amino acid substitutions.
  The geometric mean of strain, $\langle \dNi \rangle$, as a function of $\dm$ shows that
  PDB and AF-predicted structures permit reliable estimation of mutation effects up to about \SI{15}{\angstrom} from mutated residues (\fref{fig:fig6}B); at high $\dm$, strain increases since extremal residues tend to be in flexible loops and tails.
  Averaging ($\AFave$) across many AF-predicted structures enables measurement of fine-grained mutation effects up to \SI{25}{\angstrom} from mutated residues.
  Altogether, our results demonstrate shows how evolutionary perturbations (\ie, mutations) induce structural changes that can be seen as effective internal motion of proteins. The natural measure of this effective motion is strain, which decays exponentially with distance from the perturbation.

\subsection*{Evolutionary deformation is linked to function.}
  We have shown that evolutionary perturbations give rise to changes in structure, akin to motion due to physical perturbations.
  Ideally, we would discuss the effect of evolutionary perturbations on dynamics as well, but this is less clear -- recent augmentations to the AlphaFold algorithm may enable such investigations, but it is yet to be seen whether AlphaFold can inform the effect of mutations on dynamics.~\cite{delalamoSampling2022,wayment-steelePredicting2023} Nevertheless, we ask whether one can deduce how a mutation affects function merely from knowing the structural change such a perturbation induces.
  
  To address this question, we studied three protein families and examined the relationship between evolutionary strain
  and phenotype for proteins whose genotype-to-phenotype space has been partially mapped via deep mutational
  scanning experiments.~\cite{mcbrideAlphaFold22022} Here we highlight one such protein, the blue fluorescent
  protein (BFP, \fref{fig:fig7}C).~\cite{poenc19} We find that by measuring the strain induced at the chromophore binding site
  by mutations at a range of sites, we can predict with great accuracy
  the effect of these mutations on fluorescence (\fref{fig:fig7}C-D). Fluorescence dims if strain at the chromophore binding site exceeds \num{0.008}, and if it passes exceeds \num{0.013} then
  fluorescence is lost entirely. 
  
  Our findings shine light on the elaborate mapping between sequence and function -- that allosteric deformations at a functionally important residue can be sufficient to explain much of the variance in function.
  Furthermore, we have shown that AF-predicted strain correlates with other important phenotypes, such as catalytic activity and stability.~\cite{mcbrideAlphaFold22022,mcbrideAlphaFoldpredicted2023}
  These results demonstrate how deep mutational scanning, along with AlphaFold, have opened up a scintillating new avenue for understanding how proteins work and evolve.

\section{Discussion}
  We have presented a theory of why motion is essential for protein function, and how it is encoded in structure and sequence.
  Although there are multiple hypotheses about why protein motion is important for function, we have proposed that it is absolutely essential for achieving \emph{superlative discrimination} between molecules.
  This motion is encoded throughout proteins, due to the long-ranged nature of force transduction in elastic (and plastic) networks.
  In the following, we will highlight several predictions that emerge from our physical and mechano-chemical models of proteins, and suggest experimental tests for these predictions.
  
\subsection*{Binding-induced strain highlights functional residues.}
  Specific binding and allosteric force transduction is achieved by a precise calibration of the dynamical response to physical interactions between proteins and ligands (\fref{fig:fig6}).
  The residues that contribute most to this dynamical response are those that undergo high strain upon binding.~\cite{ravasioMechanics2019}
  These are exactly residues that have an outsize effect on the effective spring constant $\Keff$ that dictates the internal motion of the protein in response to binding. In particular, the existence of high-strain regions is found to be crucial for long-range transduction of force (\fref{fig:fig6}).
  We show that the appropriate metric for measuring deformation is strain, and that computing the effective strain is a simple method for reliably estimating deformation (\fref{fig:fig5}).

  What follows is a potential answer to the longstanding question of how residues \emph{far} from binding sites affect the binding.
  We predict that mutations at residues that experience high strain upon binding should lead to large changes in function, even if they are far from the binding site.
  At first, one might think that the structures in the PDB offer a quick way to test this hypothesis. However, PDB data should be used with great caution since strain measured using PDB structures is often a sign of thermal fluctuations in flexible regions~\cite{mcbrideAlphaFold22022}, although the strain may still be informative in cases of significant motions.
  Instead, based on our experience of using AF, we recommend measuring strain between conformational ensembles, by averaging over multiple structures to arrive at a more precise measure of functional deformation.~\cite{mcbrideAlphaFold22022,mcbrideAlphaFoldpredicted2023}
  Thus, strain measurements can be used to target mutations far from the binding site that are predicted to have a significant effect on catalysis and/or binding, which can be measured experimentally.

\subsection*{Possible links of physical strain and evolutionary strain.}
  Theoretical results from elastic network models of force propagation,~\cite{tlustyPhysical2017,duttaGreen2018,eckmannColloquium2019,mcbrideGeneral2022}
  along with empirical and computational estimation of effects of mutations
  with distance from mutated sites,~\cite{mcbrideAlphaFold22022,mcbrideAlphaFoldpredicted2023} show that force can propagate throughout
  networks of connected residues. It stands to reason that the same
  underlying physics can explain both phenomena. In order to show an equivalence between evolutionary and physical perturbations, we can study strain in response to both force and mutation. We can also study free energy landscapes 
  before and after mutation, and before and after binding. Is there a rigorous equivalence, even partially, between the effects of binding and mutation? Answering this would yield a unified theory of proteins where physics and evolution are intertwined. 

\subsection*{Towards a phase diagram of specificity for real proteins.}
  We presented a theoretical affinity-specificity phase diagram for
  a minimal model of protein-ligand binding.~\cite{mcbrideGeneral2022} This study can help conceptualize the factors leading to successful molecular discrimination and their interplay. However, by studying a thermodynamic
  model using an elastic network, we ignore two important factors:
  binding kinetics and plastic deformation.
  To truly understand how binding specificity works, we need precise measurements of real proteins. One benefit of the minimal model that we would like to retain is the simplicity of the specificity phase diagram, which has only three dimensions: mismatch, flexibility, and energy level.
  Furthermore, we demonstrated that the dimension can be even further reduced, where the optimal mismatch is a 1D function of $\epsilon/K$. To achieve similar simplicity in real proteins is difficult as we need to know how many dimensions are required to characterize specificity and what one should measure.

  Creating an affinity-specificity phase diagram for real proteins is challenging because of the sheer number of required measurements of multiple proteins: protein-ligand mismatch, bond energy, flexibility, affinity, and specificity.
  Shape mismatch is conceptually simple but a clear metric for estimating shape mismatch between arbitrary 3-dimensional shapes is lacking (although advances in machine learning may make this possible).~\cite{simonovskyDeeplyTough2020}
  As an alternative, we suggest estimating shape mismatch indirectly, by measuring instead how much a protein needs to deform upon binding. This can be estimated using strain and studied using experimental structural data~\cite{wangPDBbind2005}, machine learning predictors,~\cite{tubianaScanNet2022,hekkelmanAlphaFill2023}
  flexible docking,~\cite{fanProgress2019} or MD simulations.~\cite{siebenmorgenComputational2020,wanRapid2020}
  Bond energy can be calculated using a similar approach, and is 1D by definition. In contrast, flexibility is hard to reduce to a single dimension since the binding mode must be identified, and when ligands are large molecules (\eg, proteins, nucleotides), or when there are multiple simultaneously-binding subtrates,
  there may be several binding modes.~\cite{orellanaLargeScale2019} But at least in the case of small molecules, it may be possible to represent a single binding mode in terms of one effective spring constant $\Keff$.
  This can be measured using MD simulations, or elastic network models.
  Finally, unlike the above quantities, affinity and specificity can be measured by existing fast and accurate experimental assays.
  Additionally, such measurements are readily available in datasets,~\cite{gilsonBindingDB2016,jankauskaiteSKEMPI2019,lingePLBD2023} or published alongside high-throughput experiments.~\cite{adics21}
  Computational tools such as MD~\cite{deruiterAdvances2020} and machine learning~\cite{jonesImproved2021} can also be used to predict affinity.
  Thus we expect that, although it requires much work, the determination of an affinity-specificity phase diagram is well within reach using existing data and methods. This direction may open a new era of theory in protein science, where theories of binding are quantitative and predictive,~\cite{shoemakerSpeeding2000,mcbrideGeneral2022} rather than qualitative and descriptive.~\cite{koshlandApplication1958,koshlandComparison1966,monodNature1965}

%\nsb{Linking binding mechanisms and specificity predictions.}
  %NEW IDEA: Correspondence between \fref{fig:fig2} and \fref{fig:fig3}C. Conformational change upon binding is equivalent to shape mismatch, while degree of conformational fluctuations is equivalent to flexibility. The \textbf{MeCh} model predicts that for a given pair of ligands, proteins with highest selectivity will have a similar ratio of flexibility to shape mismatch.
  %In terms of binding mechanism, the proteins optimized for selectivity will exhibit a proportional the degree of induced fit and conformational selection.
 % This can be tested by... check papers to see how people distinguish between the two mechanisms, see if there is something there!
%\cite{marshProbing2012}

\subsection*{Does molecular discrimination limit protein size?}
  The \textit{size-specificity-scaling} hypothesis predicts that
  specificity requirements impose a minimum size constraint on proteins (\fref{fig:fig4}A). The most
  obvious way to test this would be to take a protein and to try to make it smaller without decreasing its functional specificity. 
  However, this is extremely impractical due to the severely deleterious impact of deletions.~\cite{topolskaDeep2023}
  Inverting this approach seems considerably more tractable: take proteins of different size that are known to bind to the same molecule, and try
  to engineer mutants using directed evolution that maximize specificity.
  Amino-acid binding proteins are a promising test case for this approach since there are many of these, ranging in size from hundreds to a thousand amino acids (\fref{fig:fig4}C).
  Our hypothesis has two clear predictions: larger proteins will be able to achieve higher selectivity, and they will also be easier to evolve.

  \subsection*{Acknowledgements}
    This work was supported by the Institute for Basic Science, Project Code IBS-R020-D1.\\

\bibliography{protphys}

% \clearpage
% % {\small
% \noindent\sb{Supplementary Materials}\\

\end{document}

\title{
{\smaller[0.5] Supplemental Material for} \\ ``AlphaFold-predicted deformation from wild-type protein correlates with changes in stability''}

  \author{John M. McBride}
    \email{jmmcbride@protonmail.com}
    \affiliation{Center for Soft and Living Matter, Institute for Basic Science, Ulsan 44919, South Korea}
  \author{Tsvi Tlusty}
    \email{tsvitlusty@gmail.com}
    \affiliation{Center for Soft and Living Matter, Institute for Basic Science, Ulsan 44919, South Korea}
    \affiliation{Departments of Physics and Chemistry, Ulsan National Institute of Science and Technology, Ulsan 44919, South Korea}

\maketitle
\tableofcontents

\section{PDB Structure Data}
\subsection{How many structures are needed to average?}
  Since we found that correlations between $\dNn$ and $\ddG$ are
  higher when using averaged AF-predicted structures, a pertinent
  practical question is, how many structures should one use?
  We looked at the correlations (both $\ddG$ and $|\ddG|$) as a
  function of the number of structures used to create average
  structures, finding that most of the improvement is seen when
  using only five structures instead of one. We suggest that using
  \num{20} structures is sufficient to achieve good performance.

\subsection{pLDDT is less informative of stability than strain.}
  Since AF's predicted confidence measure, pLDDT, is an indicator
  of disorder, we also investigated whether changes in pLDDT
  are informative of changes in stability. We compared changes in pLDDT
  at the mutated site $m$, changes in the averaage pLDDT (across the entire protein),
  for both single and averaged structures, against both $\ddG$ and $|\ddG|$.
  We find poor correlations between changes in pLDDT and $\ddG$, in agreement
  with \citet{pakpo23}.

\subsection{Insufficient PDB data to measure empirical strain-stability correlation.}
  We sought to repeat this analysis using experimental structures from 
  the Protein Data Bank (PDB). We only found $\sim 100$ entries which
  have both WT and mutant experimental structures. We removed any
  NMR structures, and only considered pairs with the same oligomeric state
  and same bound substrates. This left us with \num{36} pairs of PDB
  structures, which is too few to study statistical correlations.
  It is possible that there are sufficient $\ddG$ measurements
  across other data sets, but the limit of unique mutations
  is determined by the number of possible structure pairs that
  differ by a single amini acid substitution. In a previous study \cite{mcbbi23}
  we only found $\sim$\num{1000} of these, so there is some hope
  that this analysis can be performed is the other datasets have
  enough $\ddG$ measurements.

\subsection{Construction of a reduced ThermoMutDB set}
  Our original approach was to use the list of mutations data provided by \citet{pakpo23}.
  We decided to work exclusively with the original ThermoMutDB data, for a number
  of reasons. We found that the ``TMDB\_ID'' values do not always match with the original
  ThermoMutDB ID numbers. We also found a case where a mutant (P00644, L89V)
  was assigned an incorrect mutation position; in this case, 
  both the PDB SEQRES and Uniprot sequences have a lysine at position 89, but
  the correct mutation code should be L171V.

  We found several issues in the ThermoMutDB that led to exclusion or correction
  of entries. We found that for some entries, the reference protein for which the mutation
  was defined (and $\ddG$ was calculated) was not the WT sequence found in Uniprot;~\cite{unina18,merst99,danpn08,danpn10}
  we excluded these.
  We found duplications of entries, where measurements from one paper were referenced
  in several other papers, yet both were included;~\cite{robjp18} we excluded these entries.
  We found that one measurement (Y27A) in one source,~\cite{shobi90} and all measurements in another
  source~\cite{carbi95} were included with the incorrect sign; we corrected these.

  \begin{figure*}[t!]  \centering
  \includegraphics[width=0.99\linewidth,]{Figures/si1.pdf}
  \caption{\label{fig:si1}
  \textbf{AF predictions are more variable in regions with low pLDDT.}
  Strain per residue $\dNi$ vs. pLDDT per residue for:
  (A) Strain calculated between repeat AF predictions of the same protein sequence;
  blue dots are shown for different pairs of single structures; black circles
  indicate means across multiple pairs.
  (B) Strain calculated for pairs of averaged structures, whose sequences differ
  by a single mutation; residues with pLDDT $<70$ are excluded.
  Pearson's correlation coefficient and p value are shown.
}
  \end{figure*}

  \begin{figure*}[t!]  \centering
  \includegraphics[width=0.90\linewidth,]{Figures/si2.pdf}
  \caption{\label{fig:si2}
  \textbf{Improving precision of mutation effect prediction.}
  Strain at the mutated site $m$, $\dNm$ (left), and near the
  mutated site, $\dNn$ (right), for four different methods of calculating
  strain: (A) All residues and pairs of single structures.
  (B) All residues and pairs of averaged structures.
  (C) Only residues with pLDDT $>70$ and pairs of single structures.
  (D) Only residues with pLDDT $>70$ and pairs of averaged structures.
  Spearman's $\rho$ and p value are shown.
}
  \end{figure*}

  \begin{figure*}[t!]  \centering
  \includegraphics[width=0.99\linewidth,]{Figures/si3.pdf}
  \caption{\label{fig:si3}
  \textbf{Biased sampling of IgG-binding protein G (IgGb).}
  A: pLDDT vs sequence position $i$ for IgGb.
  B: Distribution of mutations in the ThermoMutDB for IgGb along
  the sequence, and the corresponding $\ddG$ values.
}
  \end{figure*}

  \begin{figure*}[t!]  \centering
  \includegraphics[width=0.99\linewidth,]{Figures/si4.pdf}
  \caption{\label{fig:si4}
  \textbf{Buried residues tend to be within \SI{15}{\angstrom} of other resiudes.}
  For each residue $i$, its neighbors are defined as the set of residues whose
  $\CA$ distances are within \SI{13}{\angstrom}. Plot shows average number of neighbors of residues
  within $\gamma$ of the mutated site $m$, as a function of $\gamma$.
}
  \end{figure*}

  \begin{figure*}[t!]  \centering
  \includegraphics[width=0.99\linewidth,]{Figures/si5.pdf}
  \caption{\label{fig:si5}
  \textbf{$\dNn$--$\ddG$ correlations calculated using FoldX structures.}
  Strain near mutated site $\dNn$ vs. $\ddG$ (A) and magnitude of stability
  change $|\ddG|$ for all mutations in our reduced ThermoMutDB sample
  of \num{2500} unique mutants. Spearman's $\rho$ and p value are shown.
}
  \end{figure*}

  \begin{figure*}[t!]  \centering
  \includegraphics[width=0.99\linewidth,]{Figures/si6.pdf}
  \caption{\label{fig:si6}
  \textbf{Correlation between distance from mutated site $\delta_m$ and $\dNi$}
  for all proteins in our reduced ThermoMutDB sample.
  Different distributions are shown for shear calculated using:
  AF-predicted structures, all residues and pairs of single structures (AF).
  AF-predicted structures, only residues with pLDDT $>70$ and pairs of single structures (AF; pLDDT $> 70$).
  AF-predicted structures, all residues and pairs of averaged structures (AF; Ave).
  AF-predicted structures, only residues with pLDDT $>70$ and pairs of averaged structures (AF; pLDDT $> 70$; Ave).
  FoldX-predicted structures, all residues and pairs of single structures (FoldX).
  FoldX-predicted structures, all residues and pairs of averaged structures (FoldX; Ave).
}
  \end{figure*}

  \begin{figure*}[t!]  \centering
  \includegraphics[width=0.99\linewidth,]{Figures/si7.pdf}
  \caption{\label{fig:si7}
  \textbf{$\Delta$pLDDT is a poor predictor of $\ddG$.}
  Spearman's correlation between differences in pLDDT and stability.
  Differences in pLDDT are measured in four ways:
  difference in pLDDT at the mutated site $\dP_m$;
  average difference in pLDDT across the whole protein $\langle \dP_i \rangle$;
  difference in pLDDT at the mutated site, averaged over 50 pLDDT predictions $\dP_m^{ave}$;
  average difference in pLDDT across the whole protein,
  averaged over 50 pLDDT predictions $\langle \dP_i \rangle^{ave}$;
}
  \end{figure*}

  \begin{figure*}[t!]  \centering
  \includegraphics[width=0.99\linewidth,]{Figures/si8.pdf}
  \caption{\label{fig:si8}
  \textbf{Effect of the number of structures used to create averages.}
  Strain-stability correlation for all mutations in our reduced ThermoMutDB sample
  of \num{2500} unique mutants, as a function of the number of structures used
  to create average structures.
}
  \end{figure*}

\clearpage

\bibliography{af2_stab}